\documentclass[12pt]{article}
\newcommand{\ba}{\noindent $\begin{array}}
\newcommand{\ea}{\end{array}$}
\newcommand{\be}{\begin{equation}}
\newcommand{\ee}{\end{equation}}
\newcommand{\bd}{\begin{displaymath}}
\newcommand{\ed}{\end{displaymath}}
\newcommand{\beq}{\begin{eqnarray*}}
\newcommand{\eeq}{\end{eqnarray*}}
\newcommand{\beqn}{\begin{eqnarray}}
\newcommand{\eeqn}{\end{eqnarray}}
\usepackage{amsfonts,latexsym,lscape}

\small\normalsize

\usepackage{rawfonts}
\newtheorem{theorem}{Theorem}[section]

\newfont{\Bb}{msbm10 scaled\magstep1}

\begin{document}

\pagestyle{plain} \thispagestyle{empty}
\vspace{1.5cm}

\begin{center}
{\large \textbf{AN ABS ALGORITHM FOR A CLASS OF SYSTEMS OF
STOCHASTIC LINEAR EQUATIONS}}
\end{center}

\begin{center}
{\textbf{
                  Hai-Shan Han\footnote{CORA, Department of Applied Mathematics,
                               Dalian University of Technology
                               Dalian 116024, China; Department of Mathematics,Inner Mongolia University of Nationalities,Tongliao 028000 China},
                  Antonino Del Popolo\footnote{Dept. of Mathematics, University of Bergamo,
                               via dei Caniana 2, 24127 Bergamo, Italy; \\
Istanbul Technical University, Ayazaga Campus, Faculty of Science and Letters, 34469 Maslak/Istanbul, Turkey} and
                 Zun-Quan Xia\footnote{CORA, Department of Applied Mathematics,
                               Dalian University of Technology
                               Dalian 116024, China, zqxiazhh@dlut.edu.cn}
                }
}
\end{center}

\vspace{2mm} \vspace{2mm}
\begin{center}
\parbox{13.5cm}{\small
\textbf{Abstract.}  This paper is to explore a model of the ABS
Algorithms dealing with the solution of a class of systems of linear
stochastic equations $A\xi=\eta$ when $\eta$ is a $m$-dimensional normal distribution. 
It is
shown that the stepsize $\alpha_i$ is distributed as $N(u_i,\sigma_i)$ (being $u_i$
the expected value of $\alpha_i$ 
and $\sigma_i$ its variance) and the approximation to the solutions $\xi_{i}$ is distributed as $N_n(U_i,\Sigma_i)$ (being $U_i$
the expected value of $\xi_i$ 
and $\Sigma_i$ its variance) , for this algorithm model.
\\[10pt]
                           }
\end{center}

\section{Introduction}
Since the early fifties, the theory of stochastic equations, and in particular
stochastic integral equations, has been the stage for intense activity. 
This has been stimulated in part by an immense wealth of applications beginning with the
Langevin equation in statistical mechanics and Wiener's emphasis on the role
of randomness in the problems of cybernetics. Applications of random equations
are now found in many areas of engineering, physical and biological sciences
and systems theory. 

%

Historically the study of random equations (stochastic integral equations, stochastic operator equations, stochastic difference equations,  etc.) 
started with that of integral equations.
The theory of the last 
has grown along two parallel branches. 
In one, initiated by K. Ito in 1951, the source of randomness is a white noise term leading to an important 
class of stochastic integral equations which has developed hand in hand with the theory of Markov
processes. In the other, classical linear and nonlinear integral equations with
random right-hand sides, random kernels or defined on random domains are
studied. 


The application of stochastic equations is more extensive than deterministic equations, because random factors
are usually included in many natural phenomena.

The study of random linear equations can at least be traced back to the works of [Bellman 54], [Rosenbloom 55],  [Kalman 62].
Random linear difference equations arises naturally in many areas of engineering sciences 
including automatic control and signal processing, and have drawn much attention from researchers in the
past several decades [Brandt 86].

Attempts were made by many scientists and mathematicians to develop and unify the theory of random equations employing
concepts and methods of probability theory and functional analysis. The Prague
School of probabilists under Spacek initiated a systematic study using probabilistic
operator equations as models for various systems. This development was further energized
by the survey article by [BhRe 72] on various treatments of random
equations under the framework of functional analysis. 

Several techniques have been proposed to solve stochastic linear equations (see for instance [Brandt 1986], [Verma 1996]).
Stability problems of random linear equations
have also been studied in [Guo 90], [Guo 93], [GuoLj 95]. 

An approach that has never been tried to solve stochastic linear equations is through the use of ABS methods, introduced by Abaffy, Broyden and Spedicato in 1984 in the form of a large class of methods for solving linear determined or underdetermined systems. The original formulation of 1984 has been later generalized giving a larger class of methods that essentialy contains all possible methods building a sequence of approximations $x_i$ to the solution having the following property: for $x_1$ arbitrary, the vector $x_{m+1}$ solves the given system. ABS methods have been applied the least squares problem, to non-linear equations, to linear and non-linear programming and eigenvalue problems.

Before showing how the ABS algorithm can be used  to solve a system of stochastic equations of the form 
\begin{equation}\label{hhx2}
  A\xi(w)=\eta(w)
\end{equation}
where $\xi(w)$, $\eta(w)$ are random vectors, we recall the ABS algorithm. 

%
%

Consider the general linear system, 
\begin{equation}\label{hhx3}
    Ax=b  \hspace{0.5cm} {\rm or} \hspace{0.5cm} a_i^Tx=b_i,\quad i=1,\cdots,m
\end{equation}
where $x \in I\!\!R^m$,\quad $b \in I\!\!R^n$,\quad $m \leq n$
\quad and \quad$A \in I\!\!R^{m \times n}$, $A=(a_1, a_2,\cdots,
a_m )^T$.

The class of ABS algorithms to solve Eq. (\ref{hhx3})
was originally introduced by Abaffy, Broyden and Spedicato (1984), see [AbBs84] and [AbSp89].
The iterate  scheme of the basic ABS class of algorithm is defined as follows:

\bigskip
\noindent \textbf{Basic (unscaled) ABS Class of Algorithms}:
\,[AbBs84]\,,\,\, [AbBs89]:
\begin{description}
\item[\textbf {  (A) }]  Initialization.\\[1pt]
                     Give  an arbitrary vector $ x_1 \in  I\!\!R^n $,  and an arbitrarily nonsingular matrix $ H_1 \in R^{n,n} $. Set $ i = 1 $.
\vspace{-2mm}
\item[\textbf{  (B)}] Compute $s_i = H_i a_i$ and $\tau_i=\tau^Te_i =a_i^T x_i-b^T e_i$\\[1pt]
\vspace{-2mm}
\item[\textbf{  (C)}] Check the compitability of the system of linear equations.\\[1pt]
                    If $ s_i \not= 0 $ then goto (D). \\
                    If $ s_i = 0 $ and $ \tau_i = 0 $ then set $x_{i+1} = x_i$, $H_{i+1} = H _i$
                  and goto (F), the $ i $-th equation  is a linear combination of the previous equations. Otherwise  stop, the
                  system has no solution.
\vspace{-2mm}
\item[\textbf{ (D)}] Compute the search vector $p_i\in I\!\!R^n$ by
                        $p_i=H_i^Tz_i$
                  where $z_{i}$, the  parameter of Broyden, is arbitrary
                  satisfying the condition 
                          $z_i^TH_ia_i\neq 0$
\vspace{-2mm}
\item[\textbf{ (E)}] Update the approximation of  the solution $ x_i $ by
                        $x_{i+1} = x_i - \alpha _i p_i$,
                where the stepsize $ \alpha _i $ is computed by
                        $\alpha _i = \tau_i /a_i^T p_i$
                  If $i=m$ stop; $x_{m+1}$ solves the system.
\vspace{-2mm}
\item[\textbf{ (F)}] Update the (Abaffian) matrix $ H_i. $
                   Compute
                        $H_{i+1} = H_i - H_ia_iw_i^TH_i /w_i^TH_ia_i$
                   where $ w_i \in I\!\!R^n $, the parameter of Abaffy, is arbitrary satisfying the condition
                              $$w_i^T H_i a_i = 1 \textrm{ or } \neq 0$$
\vspace{-2mm}
\item[\textbf{ (G)}] Increment the index $ i $ by one and goto (B).
\end{description}

\bigskip
\noindent We define $n$ by $i$ matrices $A_i,\, W_i$ and $ P_i$ by
$$    \begin{array}{l}
                  A_i=(a_1,\,\cdots,\,a_i)^T, \quad
                  W_i=(w_1,\,\cdots,\,w_i),\quad
                  P_i=(p_1,\,\cdots,\,p_i)

   \end{array}
$$
\bigskip
\noindent Some  properties of the above recursion (see for instance,  Abaffy and Spedicato (1989), \cite{AbSp 89}), that are the basic formulae to be used later on, are
listed below .
\begin{description}
\item[\textbf{ a. }]  Implicit factorization property
                     $A_i^TP_i = L_i$
                    with  $ L_i $ nonsingular lower triangular.
\vspace{-8pt}
\item[\textbf{ b.}]  Null space characterizations
$$
                   \begin{array}{c}
                   {\cal N} (H_{i+1}) ={\cal  R} (A_i^T),\quad
                   {\cal N} (H_{i+1}^T) ={\cal R} (W_i),\quad  \\[6pt]
                   {\cal N} (A_i) ={\cal R} (H_{i+1}^T)
                   \end{array}
$$ 
                where ${\cal N}$= Null and ${\cal R}$=Range.
\vspace{-8pt}
\item[\textbf{ c.}]\ \ The linear variety containing all solutions to $Ax=b$
                   consists of the vectors of the form
                          $x = x_{i+1} + H^T_{i+1}q$
                 where $ q \in I\!\! R^n $ is arbitrary.
\end{description}

As we shall show in the rest of the paper, this method can be used to solve a linear stochastic system in the following way.
For a linear system 
$$A \xi= \eta$$ with a $m \times n$ 
matrix of arbitrarary rank with $m \leq n$, the so-called ABS algorithm
produces a solution
$$\xi= A^I \eta,$$
with $A^I$ some generalised inverse.
Now it is quite clear that if $\eta$ is a normally distributed vector, so is $\xi$, or in other words
in the original system there are also normally distributed vector so is $\xi$ as solutions.
The mean of $\xi$ can obviuosly be computed from the mean of $\eta$ in the usual deterministic
way (save for a component in the kernel of A).
So the only interesting question left is how to compute the covariance of $\xi$ from that
of $\eta$. Formally it is $A^I V (A^I)^T$ if the covariance of $\eta$ is V . The paper gives a recursive
formula through the ABS algorithm for the special case that the components of $\eta$ are
uncorrelated, i. e. V = I.

This paper is organized as follows. In Section 2, some basic
definitions and operations that will be used below are presented. 
A basic ABS algorithm, called ABS-S for stochastic linear equations is given in Section 3. 
Some properties of the approximations to solutions and 
$\alpha_i$, are given in section 4. 
In section 5, the ABS-S algorithm is illustrated by an example.
In section 6, $\alpha_i$ is discussed and same results
are also given.

\section{Preliminaries }
\setcounter{equation}{0}

 Consider a class of systems of stochastic
 linear equations being of the form

\begin{equation}\label{hhx5}
A\xi(w)=\eta(w)
\end{equation}

In this paper, we present an ABS algorithm under the condition that
$\eta(w)$ has a $m$-dimensional normal distribution. 
It will be
shown that the iteration step length $\alpha_{i} \sim N(u_i,\sigma_i)$ 
(being $u_i$
the expected value of $\alpha_i$ and $\sigma_i$ its variance) and 
the iteration solution $\xi_{i}$ is $N_n(U_i,\Sigma_i)$, $i$=1,2,...$m$+1.


We show that 
$(u_i,\sigma_i)$ and $(U_i,\Sigma_i) $ are determined by
four iterative formulae, that $\xi_{i+1}$
is a solution of the first $i$ equations and we discuss the step length 
$\alpha_{i}$.
~\\

{\bf Definition 2.1}[GoMi84] Let ($\Omega,\cal A, \cal P$) be a probability space and 
let $\xi$: $\Omega \rightarrow R$ be a real-valued function on $\Omega$. We say that $\xi$ is a random variable on $\Omega$ iff
for every $x \in R$, $\{\omega: \xi(\omega) \leq x\} \in \cal A$.
~\\

{\bf Definition 2.2} If a random variable $\xi$ has a normal distribution with expectation $\mu$ and variance $\sigma^2$, then it is denoted by 
$\xi \sim N(\mu,\sigma^2)$.
~\\

{\bf Definition 2.3}[GoMi84] $\eta=(\eta_1,\eta_2,....,\eta_m)$ has a joint $N(\mu, \Sigma)$ density iff there exists an $m \times n$ matrix $A(m \leq n)$ of rank $m$ such that $\eta = A \xi +\mu$ and $\xi=(\xi_1,\xi_2,....,\xi_n)$ is a vector of independent N(0,1) random variables. Equivalently $\xi$ is a $N(0,I_n)$ multiple normal density.
~\\

{\bf Definition 2.4} A random variable of random distribution $\xi$ is said to be equal to the random variable of random distribution $\eta$, denoted by $\xi=\eta$, if the distributions of $\xi$ and $\eta$ are the same and E$\xi$=E$\eta$ and D$\xi$=D$\eta$, where E and D stands for expectation value and variance, respectively.
~\\

{\bf Definition 2.5} Consider a system of random linear equations A$\xi$=$\eta$. Suppose that $\eta$ has a certain distribution. If the distribution of $\xi$, denoted by $\xi_0$, is given, then $\xi_0$ is a solution of equations A$\xi$=$\eta$.
~\\

{\bf Property 2.1}[GoMi84] If $\xi=\eta+a$, $a \in  \cal R$, then E$\xi$=E$\eta$+a, D$\xi$=D$\eta$.
~\\

{\bf Property 2.2}[Ande 84] If ($\xi,\eta$) is a 2-dimensional normal distribution, then $a \xi+b\eta$ is also a normal distribution, where $a$ and $b$ are constants.

\section{ABS-S Algorithm to solve System of Stochastic Linear Equations}
\setcounter{equation}{0}

\noindent
Consider a system of stochastic linear equations
\begin{equation}\label{hhx8}
    A\xi=\eta
\end{equation}
where $\eta=(\eta_1,\eta_2,\cdots,\eta_m)^T$ is a $m$-dimensional
stochastic vector and $A=(a_1,a_2,\cdots,a_m)^T \in R^{m,n}$

\noindent
 The ABS-S algorithm, based on the basic ABS algorithm, is defined, 
 as follows.
\bigskip \noindent
 {\bf  ABS-S Algorithm to solve systems of stochastic linear equations}
\begin{description}
\item[\textbf {(A1) }]  Initialization.\\[1pt]
                     Give  an arbitrary stochastic vector $ \xi_1 \in  I\!\!R^n $,  and an arbitrarily nonsingular matrix $ H_1 \in R^{n,n} $.
                     Set $ i = 1 $ 
\item[\textbf{(B1)}] Compute the two quantities:
                     \\[1pt]
                      
                       $$
                          \begin{array}{l}
                                   s_i = H_i a_i, 
                                   \tau_i=a_i^T \xi_i- \eta_i
                          \end{array}
                       $$
\item[\textbf{  (C1)}] Check the compatibility of the system of linear equations.\\[1pt]
                    If $ s_i \not= 0 $ then goto (D). \\
                    If $ s_i = 0 $ and $Prob.( \tau_i = 0)=1 $ then set
                      $$
                                    \begin{array}{l}
                                         \xi_{i+1} = \xi_i \\
                                         H_{i+1} = H _i
                                  \end{array}
                     $$
                  and go to (F), the $ i $-th equation  is a linear combination of the previous equations. Otherwise  stop, the
                  system has no solution.
\item[\textbf{ (D1)}] Compute the search vector $p_i\in I\!\!R^n$
                      by
                        $p_i=H_i^Tz_i$
                  where $z_{i}$, the  parameter of Broyden, is arbitrary
                  satisfying the condition
                          $$z_i^T H_ia_i\neq 0$$

\item[\textbf{ (E1)}] Update the random approximation $ \xi_i $ of a
solution 
by
                        $$\xi_{i+1} = \xi_i - \alpha _ip_i$$
                where the stepsize $ \alpha _i $ is computed by
                        $$\alpha _i = \displaystyle\frac{\tau_i }{a_i^Tp_i}$$
                  If $i=m$ stop; $\xi_{m+1}$ solves the system(\label(\ref{hhx8}).
\item[\textbf{ (F1)}] Update the ( Abaffian ) matrix $ H_i. $
                   Compute
                        $$H_{i+1} = H_i - \displaystyle\frac{H_i a_i^T w_i H_i
                        }{w_i^T H_i a_i}$$
                   where $ w_i \in I\!\!R^n $, the parameter of Abaffy, is arbitrary satisfying the condition
                              $$w_i^T H_i a_i = 1 \textrm{ or } \neq 0$$
\item[\textbf{ (G1)}] Increment the index $ i $ by one and go to
                     (B1).
\end{description}

\section {The properties of $\alpha_i$ and $\xi_i$}
Let $\eta \sim N_m(v,I_m)$ where $\eta=(\eta_1,\eta_2,.....,\eta_m)^T$ and $v=(v_1,v_2,....,v_m)^T$
\bigskip
\noindent
~\\

{\bf Proposition 4.1}

  $$
  \tau_1= a_1^T\xi_1-\eta_1 \sim N(a_1^T \xi_1-v_1,1)
  $$
 $$
 \alpha_1=\displaystyle\frac{\tau_1}{a_1^Tp_1}
           \sim N(\displaystyle\frac{a_1^T\xi_1-v_1}{a_1^T
           p_1},\displaystyle\frac{1}{(a_1^T p_1)^2})
 $$
\bigskip
\noindent
 \textbf{Proof.}\quad From  $\eta \sim N_m(v,I_m)$, we know that $\eta_i \sim
N(v_i,1),i=1,2,\ldots,m$. 
By property 2.2, $\tau_1$ has a normal distribution and 
  $\eta \sim N(v,I_m)$, $\eta_i \sim N(v_i,1),i=1,2,...,m$
 \begin{eqnarray*}
 E\tau_1&=& E(a_1^T\xi_1-\eta_1)=a_1^T  \xi_1-v_1;
\nonumber \\
 D\tau_1&=& D(a_1^T \xi_1-\eta_1)=1;
\nonumber \\ 
u_1=E\alpha_1&=&E((a_1^T p_1)^{-1} \tau_1) =(a_1^T p_1)^{-1}(a_1^T \xi_1-v_1); 
\nonumber \\
\sigma_1=D\alpha_1&=&D((a_1^T p_1)^{-1} \tau_1)=(a_1^T p_1)^{-2}.
\end{eqnarray*}

Therefore
\begin{eqnarray*}
  \alpha_1&=&(a_1^T p_1)^{-1} \tau_1 \sim N( u_1, \sigma_1)
\nonumber\\
 \tau_1&=& a_1^T  \xi_1-\eta_1 \sim N(E\tau_1,D\tau_1)
\end{eqnarray*}
The proof is completed. 

{\bf Proposition 4.2} If $\xi_2$ is generated by the ABS-S algorithm, then 

  $$
  \xi_2 \sim N_n(\xi_1-\displaystyle\frac{a_1^T \xi_1-v_1}{a_1^T p_1} p_1,
  \displaystyle\frac{p_1  p_1^T}{(a_1^T p_1)^2})
  $$
\bigskip
\noindent\textbf{Proof.}\quad
 By using 
Proposition 4.1, we have that $\alpha_1 \sim N(E\alpha_1,D\alpha_1)$. By property 2.2, since $\xi_2=\xi_1-\alpha_1 p_1$ has a normal distribution and

\begin{eqnarray*}
U_2=E\xi_2&=&E(\xi_1-\alpha_1 p_1)=E(\xi_1)-E(\alpha_1 p_1)=\xi_1-(a_1^T p_1)^{-1}(a_1^T \xi_1-v_1)p_1
\nonumber \\
\Sigma_2=D\xi_2&=&D(\xi_1-\alpha_1 p_1)=(a_1^T p_1)^{-2}p_1p_1^T
\end{eqnarray*}
it follows that 
$$\xi_2 \sim N(U_2,\Sigma_2)$$

   The proof is completed.

{\bf Proposition 4.3}
$ a_i p_j=0,\quad i<j$

\bigskip
\noindent
 \textbf{Proof.} see \cite{AbBs 84}, \cite{AbSp 89} 
~\\
{\bf Proposition 4.4} If $\tau_i$ is generated by the ABS-S algorithm, then $\tau_i$ has a normal distribution, and

   $$
    \tau_i \sim N(a_i^T\xi_1-a_i^T\Sigma_{j=1}^{i-1}(E\alpha_j)p_j-v_i,1+a_i^T
      \Sigma_{j,k=1}^{i-1} cov(\alpha_j p_j,\alpha_k p_k)a_i),i \geq 2
  $$
where 
$$cov(\alpha_j p_j,\alpha_k p_k)= \rho_{jk}p_j^T p_j p_k^T p_k D(\alpha_j) D(\alpha_k)$$
where $\rho_{jk}$ is the correlation coefficient of $\alpha_j p_j$ and $\alpha_k p_k$
~\\

\bigskip
\noindent
 \textbf{Proof.}\quad By the ABS-S algorithm, we have that  $ \tau_i=a_i^T \xi_i-\eta_i$,  $ \xi_{i+1}=\xi_i-\alpha_i  p_i$
 therefore
$$  \tau_i=a_i^T \xi_i-\eta_i=a_i^T ( \xi_{i-1}-\alpha_{i-1} p_{i-1})-\eta_i=...=a_i^T (\xi_1-\Sigma_{j=1}^{i-1}(\alpha_j  p_j))-\eta_i$$
Because $\alpha_j$, $j <i$ has a normal distribution, property 2.2, $\eta_i \sim N(v_i,1)$ hence $\tau_i$ has a normal
distribution, and

\begin{eqnarray*}
 E\tau_i&=&E(a_i^T (\xi_1-\Sigma_{j=1}^{i-1}(\alpha_j p_j))-\eta_i)=a_i^T  (\xi_1-\Sigma_{j=1}^{i-1}(E\alpha_j)p_j)-v_i
\nonumber \\
D\tau_i&=&D(a_i^T (\xi_1-\Sigma_{j=1}^{i-1}(\alpha_j p_j))-\eta_i)=a_i^T  [\Sigma_{j,k = 1}^{i-1}cov(\alpha_j  p_j,\alpha_k p_k)]
      a_i+1.
\end{eqnarray*}
The proof is completed.
~\\

{\bf Proposition 4.5} If $\alpha_i$ is generated by the ABS-S algorithm, then
  $$
  \alpha_i \sim N(\displaystyle\frac{a_i^T\xi_1-a_i^T\Sigma_{j=1}^{i-1}(E\alpha_j)p_j-v_i}
       {a_i^T p_i},\displaystyle
       \frac{a_i^T[\Sigma_{j,k = 1}^{i-1}cov(\alpha_j  p_j,\alpha_k p_k)]
       a_i+1}{(a_i^T p_i)^2}),i \geq 2
  $$

\noindent\textbf{Proof.}\quad
Since $ \alpha_i=(a_i^T p_i)^{-1} \tau_i$
and property 2.2, we have that the proposition holds.\\
The proof is completed. 

\begin{theorem}\label{hhx14} If $\xi_{i+1}$ is generated by the ABS-S algorithm, then 
  $$
  \xi_{i+1} \sim N_n(\xi_1-\Sigma_{j=1}^{j=i}(E\alpha_j) p_j,\Sigma_{j,k=1}^i
  cov(\alpha_j p_j,\alpha_k p_k))
  $$
 or
 $$
 \xi_{i+1} \sim N_n(E\xi_i-(E\alpha_i) p_i,D\xi_i+p_i
   D(\alpha_i) p_i^T-2 cov(\xi_i,\alpha_i p_i)), i \geq 2
 $$
 \end{theorem}

\noindent\textbf{Proof.} Since 
     $\xi_i=\xi_{i-1}-\alpha_{i-1}  p_{i-1}=\xi_{i-2}-\alpha_{i-2}  p_{i-2}-\alpha_{i-1}  p_{i-1}=\ldots=\xi_1-\Sigma_{j=1}^{j=i-1}\alpha_j p_j$,
and $\alpha_j$, $j=1,2, \ldots, i-1$ have a normal distribution, we have that $\xi_i$ has a normal distribution and
  $$E\xi_i=E(\xi_1-\Sigma_{j=1}^{i-1}(\alpha_j p_j)=\xi_1-\Sigma_{j=1}^{i-1}E(\alpha_j p_j)$$
  $$D\xi_i=D(\xi_1-\Sigma_{j=1}^{j=i-1}(\alpha_j) p_j)=\Sigma_{j,k=1}^{i-1} cov(\alpha_j p_j,\alpha_k p_k)$$
in terms of property 2.2. The proof is completed.

\begin{theorem}\label{hhx15}
   $\xi_{i+1}$ is the solution of the first $i$ equations
\end{theorem}

\bigskip
\noindent\textbf{Proof.}\quad Since $\xi_{i+1}=\xi_i-\alpha_i p_i$, one has that $a_i^T\xi_{i+1}=a_i^T(\xi_i-\alpha_i p_i)=\eta_i$. If $l < i$, then one has that 
$a_l^T \xi_{i+1}=a_l^T (\xi_i-\alpha_i p_i)=a_l^T \xi_i$. Therefore, $E(a_l^T \xi_{i+1})=E(a_l^T \xi_{i})$ and $D(a_l^T \xi_{i+1})=D(a_l^T \xi_{i})$, one has

$$E(a_l^T \xi_{i+1})=E(a_l^T \xi_{i})= \ldots=E(a_l^T \xi_l)=v_l,$$
$$D(a_l^T \xi_{i+1})=D(a_l^T \xi_{i})= \ldots=D(a_l^T \xi_l)= \ldots= D(a_1^T \xi_1)$$
The proof is completed.
~\\

{\bf Proposition 4.6}
The ABS-S algorithm terminates in finite steps.

\noindent\textbf{Proof.}\quad
The assertion comes directly from Definition (2.5), i.e. according to the definition of a solution of the system of the stochastic linear equations $A\xi=\eta$. The proof is completed.



\section{Conclusion and Discussion}
\setcounter{equation}{0}

5.1  On the step length

 Let
 $$
  E\alpha_i=\displaystyle\frac{a_i^T\xi_1-v_i}{a_i^T p_i}
 $$
 $$
 D\alpha_i=\displaystyle\frac{1+a_i^T
   \Sigma_{j,k=1}^{i-1}cov(\alpha_jp_j,\alpha_kp_k)a_i}
   {(a_i^T p_i)^2}
 $$
It follows that  
$$
    \alpha_i \sim N(E\alpha_i, \sqrt{D\alpha_i})
 $$
and hence
\begin{eqnarray}\label{hhx18}
   P(E\alpha_i-\sqrt{D\alpha_i} \leq \alpha_i \leq E\alpha_i+\sqrt{D\alpha_i})= 0.6827
\end{eqnarray}
\begin{eqnarray}\label{hhx19}
   P(E\alpha_i-2\sqrt{D\alpha_i} \leq \alpha_i \leq E\alpha_i+2\sqrt{D\alpha_i})=0.9545
\end{eqnarray}
\begin{eqnarray}\label{hhx20}
   P(E\alpha_i-3\sqrt{D\alpha_i} \leq \alpha_i \leq
   E\alpha_i+3\sqrt{D\alpha_i})=0.9973
\end{eqnarray}

Therefore, we obtain that
$$
\alpha_{i} \in [E\alpha_i-3\sqrt{D\alpha_i},E\alpha_i+3\sqrt{D\alpha_i}]
$$
in other words, 
$\alpha_i$ are determined by the initial vector $\xi_1$,
and the first component  $v_1$ of $\eta$ as well as $a_i^T p_i$.

5.2  Main results

\begin{description}
\item[1.]\quad the system of stochastic linear equations (\ref{hhx2}) with
         $\eta\sim N_m(v, I_m)$ can be solved by the ABS-S algorithm 
under some assumptions.
\item[2.]\quad The solution $\xi_i$ generated by the ABS-S algorithm has a normal distribution, if $\eta\sim N_m(v, I_m)$.
\item[3.]\quad The step length $\alpha_i$ generated by the ABS-S algorithm has a normal distribution, if $\eta\sim N_m(v, I_m)$.
\item[4.]\quad Since the matrix $A$ is non-random, $H_i,p_i$ generated by
         ABS-S algorithm are non-random, thus the ABS-S algorithm to solve
           $A\xi=\eta$ has some properties that are the same as in the case of the basic ABS algorithm.

\end{description}
5.3 Open problems
\begin{description}
  \item[1.]\quad Suppose that $\eta$ has a normal distribution, but the components of $\eta$ are dependent. An open problem is whether
the system $A\xi=\eta$ has a solution, and if it has whether it can be solved by the ABS-S algorithm.
 \item[2.]\quad Suppose that the components of $\eta$ have different distributions. Another open problem is whether the system $A\xi=\eta$ 
has a solution, and if it has whether it can be solved by the ABS-S algorithm.

\item[3.]\quad Let $A$ be a random matrix, another open problem is whether the system $A\xi=\eta$ 
has a solution, and if it has whether it can be solved by the ABS-S algorithm.

\item[4.]\quad Suppose $\eta$ is a function of a random variable.  Another open problem is whether the system $A\xi=\eta$ 
has a solution, and if it has whether it can be solved by the ABS-S algorithm.

 \end{description}

\end{document}